\documentclass[12pt]{article}

\newcommand{\AmS}{{\protect\the\textfont2
  A\kern-.1667em\lower.5ex\hbox{M}\kern-.125emS}}
\def\beq{\begin{equation}}
\def\eeq{\end{equation}}

\begin{document}
\thispagestyle{empty}
\def\thefootnote{\fnsymbol{footnote}}
\setcounter{footnote}{1}
\hfill \begin{minipage}[t]{4cm}BI-TP 99/14\\
hep-ph/9906216\end{minipage}
\vspace*{2cm}
\begin{center}
{\Large\bf Low $x_{bj}$ DIS, QCD and Generalized Vector Dominance}\footnote{
Presented at the 7th International Workshop on Deep 
Inelastic Scatterind and QCD, DESY-Zeuthen, April 19-23, 1999}
\end{center}
\vspace*{0.3cm}
\begin{center}
D.\ Schildknecht
\\[.3cm]
University of Bielefeld, Department of Theoretical Physics, \\[.3cm]
33501 Bielefeld, Germany
\end{center}
\vspace*{3cm}
\section*{Abstract}
We give a brief overview on the present status of Generalized Vector 
Dominance as appplied to vector-meson production and the total 
photoabsorption cross section in the region of small $x_{bj}$. 
We comment on how GVD originates from QCD notions such as color transparency.
\vfill

\def\thefootnote{\arabic{footnote}}
\setcounter{footnote}{0}
\clearpage

\section{THE BASIC QUESTION}

Concerning DIS at low values of the scaling variable, $x_{bj}$, a basic
question has been around for about thirty years \cite{Cornell}: when
does the virtual photon behave hadronlike, is it when $Q^2 \to 0$ or is it
when $x_{bj} \to 0$, but $Q^2$ fixed and arbitrarily large? Here,
``hadronlike'' behaviour includes the transition of the (virtual) photon
to (massive) $q \bar q$ states and their subsequent diffractive forward
scattering from the proton, in generalization of the role of the low-lying
vector mesons in photoproduction. There is qualitative experimental
evidence for this picture of generalized vector dominance (GVD)\cite{Sakurai}
at low $x_{bj}$ and large $Q^2$,
\looseness -1
\begin{itemize}
\itemsep -2pt
\item[i)] the existence of high-mass diffractive production discovered at 
HERA \cite{H1},
\item[ii)] the similarity in shape (thrust, sphericity) \cite{Zeus} of the
states diffractively produced in DIS and the ones produced in $e^+e^-$
annihilation,
\item[iii)] the persistence of shadowing in $\gamma^* A$ collisions for
$x_{bj} \to 0$ at fixed $Q^2 >> 0$ \cite{Achman}.
\end{itemize}
Quantitatively, one starts \cite{Sakurai} from the mass dispersion relation for
$\sigma_T (W^2, Q^2)$,
\beq
\sigma_T  = \int dm^2 \int dm^{\prime 2} {{\rho_T (W^2, m^2, 
m^{\prime 2}) m^2 m^{\prime 2}} \over {(Q^2 + m^2) (Q^2 + m^{\prime 2})}},
\eeq
and its generalization to the longitudinal photon absorption cross section,
$\sigma_L$, where the spectral weight function is related to the product
of the $\gamma^* q \bar q$ transition (in the initial and the final state
in the forward Compton amplitude) and the imaginary part of the $q \bar q$
proton forward scattering amplitude. Frequently, the diagonal
approximation, $\rho \sim \delta (m^{\prime 2} - m^2)$, is adopted
that requires
$\sigma_{q \bar q p} \sim 1/m^2_{q \bar q}$ to obtain scaling
for $\sigma_T$.

\section{DIAGONAL GVD}

Lack of space does not permit me to reproduce the phenomenologically
successful representation of $\sigma_{\gamma^*p} (W^2, Q^2)$ at low $x_{bj}$,
including photoproduction, by GVD. I have to refer to ref. \cite{6}.
The diagonal approximation, nevertheless, cannot be the full story.
After all, diffraction dissociation
exists in hadron reactions, and there is no particular reason in a 
gluon-exchange picture that would forbid different masses, $m_{q \bar q}
\not= m^\prime_{q \bar q}$, for ingoing and outgoing $q \bar q$ states in
the forward Compton amplitude.


\section{OFF-DIAGONAL GVD IN VECTOR-MESON PRODUCTION}
 
Reformulating and extending the off-diagonal GVD ansatz \cite{8} 
for elastic vector
meson production, recent work \cite{Schuler} by Schuler, Surrow and myself
yields a satisfactory representation of the transverse cross section
and the longitudinal-to-transverse ratio, $R$, for elastic $\rho^0, \phi$
and $J/{\rm Psi}$-production \cite{Schuler}. The theoretical 
prediction for $\sigma_{T, \gamma^*p \to Vp}$ is based on
\begin{figure}
\begin{center}
\setlength{\unitlength}{1cm}
\begin{picture}(6.5,11)
\put(-1.6,-1.2){\includegraphics{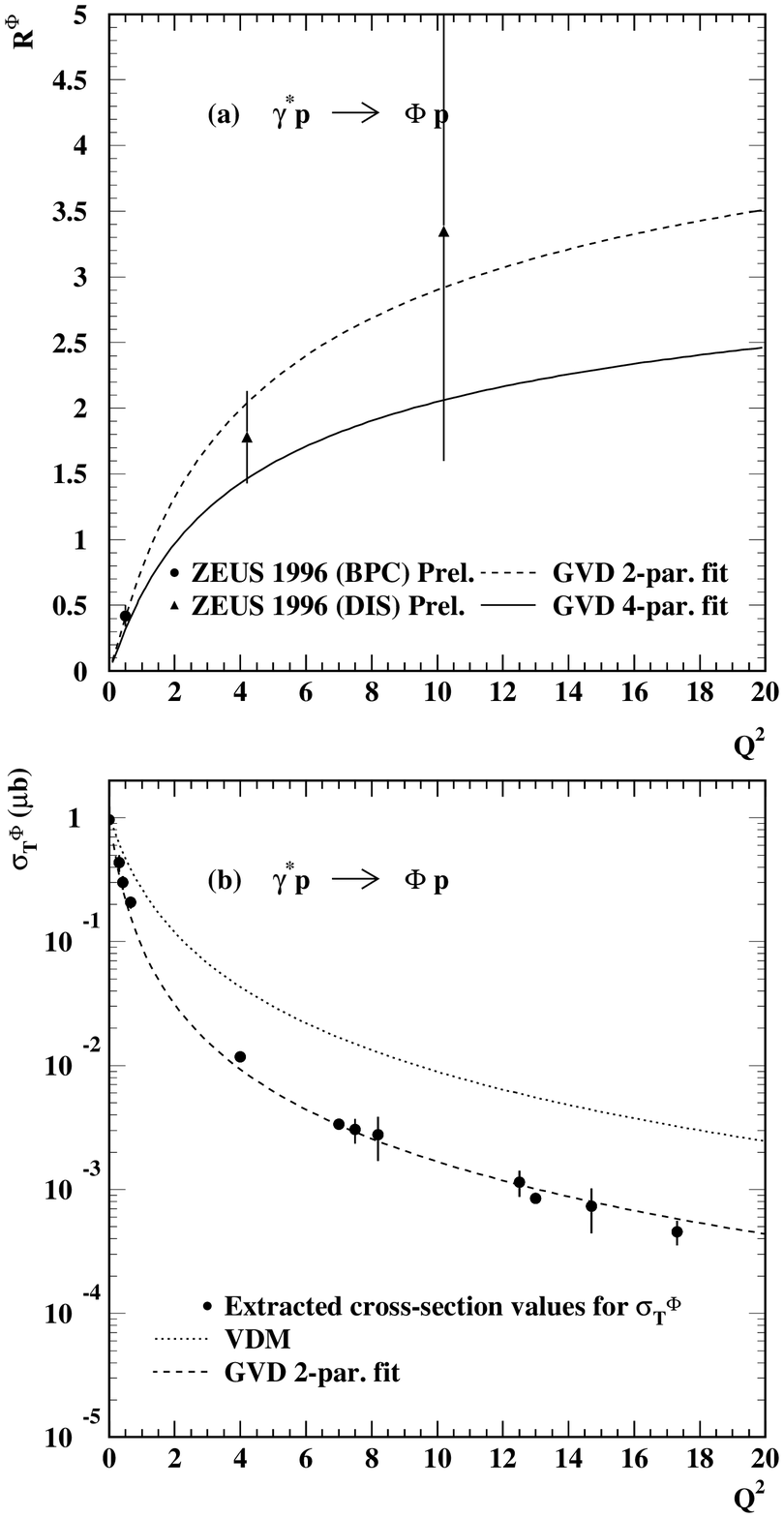}}
\end{picture} 
\end{center}
\caption{GVD in $\gamma^*p \to \phi~ p$ \cite{Schuler}.}
\end{figure}
\begin{figure}
\begin{center}
\setlength{\unitlength}{1cm}
\begin{picture}(6.5,5.0)
\put(-0.2,-2.3){\includegraphics{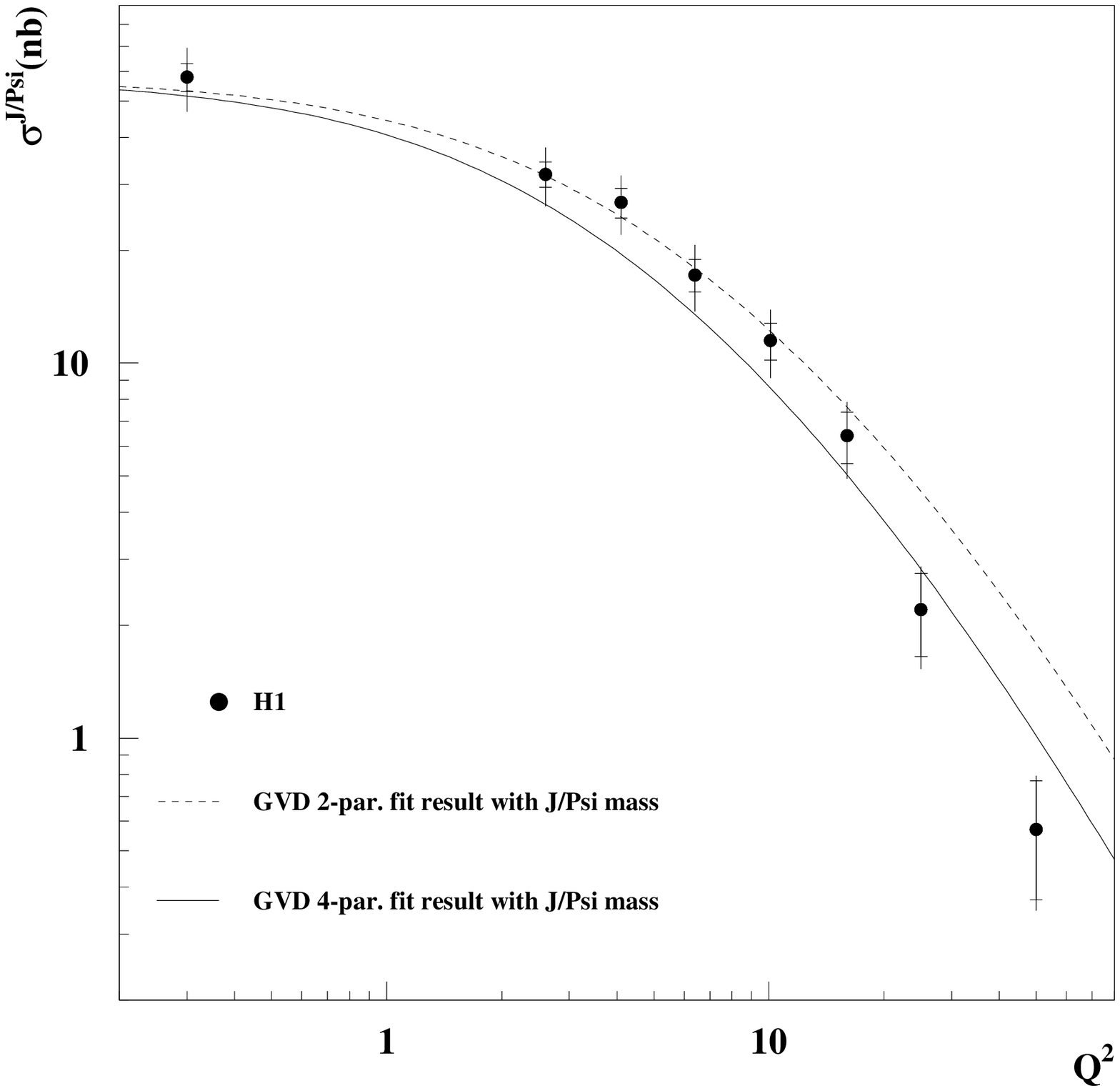}}
\end{picture} 
\end{center}
\caption{GVD in $\gamma^*p \to J/{\rm Psi}~ p$.}
\end{figure}
\beq
\sigma_{T, \gamma^* p \to V p} = {{m^4_{V,T}} \over
{(Q^2 + m^2_{V,T})^2}} \sigma_{\gamma p \to V p} (W^2).
\eeq
I refer to ref.\cite{Schuler} for the prediction for $R$. The inclusion of
off-diagonal transitions with destructive interference yields $m^2_{V,T}
< m^2_V$, where $m_V$ stands for the mass of the vector meson being
produced. As an example, in fig. 1, I show $\phi$ production. The curves
(2-par. fit) are based on $m^2_{\phi,T} = 0.40 m^2_\phi$ and 
$\sigma_{\gamma p \to \phi p} = 1.0 \mu b$. The theoretical curves for
$J/{\rm Psi}$ production in fig. 2 were obtained by the replacement $m^2_\phi
\to m^2_{J/{\rm Psi}}$ and $\sigma_{\gamma p \to \phi p} \to \sigma_{\gamma p
\to J/{\rm Psi}~ p}$.

%


\section{OFF-DIAGONAL GVD FROM QCD}

This is work in progress in collaboration with Cvetic and Shoshi \cite{Cvetic}.
Starting from the QCD notion of color transparency \cite{Nikolaev}
and an impact-parameter representation for $\sigma^{tot}_{\gamma^* p}$,
we obtain a representation for $\sigma_{\gamma^*p}^{tot}$ of the form (1).
The spectral weight function turns out to be much like the one conjectured
a long time ago \cite{12}.
Color transparency, as
fulfilled in a two-gluon exchange ansatz, provides the destructive 
interference necessary \cite{12} for convergence and scaling in (1),
thus resolving what has sometimes been 
called \cite{Bjorken} the ``Gribov paradox''.

\end{document}